\documentclass{article}

\usepackage{arxiv}

\usepackage[utf8]{inputenc} 
\usepackage[T1]{fontenc}    
\usepackage{hyperref}       
\usepackage{url}            
\usepackage{booktabs}       
\usepackage{amsfonts}       
\usepackage{nicefrac}       
\usepackage{microtype}      
\usepackage{lipsum}
\usepackage{graphicx} 
\usepackage{setspace}
\usepackage{amsmath}

\title{Phase transition of spacetime: particles as black holes in anti-de Sitter space}

\author{
  Liangsuo Shu\\
  School of Physics\\
  Huazhong University of Science \& Technology\\
  Wuhan, China \\
  \texttt{liangsuo\_shu@hust.edu.cn} \\
     \And
 Kaifeng Cui \\
  Key Laboratory of Atom Frequency Standards\\ Wuhan Institute of Physics and Mathematics\\
Chinese Academy of Sciences Wuhan, China\\
  \texttt{cuikaifeng@wipm.ac.cn} \\
     \And
Xiaokang Liu \\
 School of Energy and Power Engineering\\
    Huazhong University of Science \& Technology\\
  Wuhan, China\\
  \texttt{xk\_liu@hust.edu.cn} \\
    \And
 Wei Liu \\
  School of Energy and Power Engineering\\
    Huazhong University of Science \& Technology\\
  Wuhan, China\\
  \texttt{w\_liu@hust.edu.cn} \\
}

\begin{document}

\begin{spacing}{1.5}
\renewcommand{\baselinestretch}{1.5}
\maketitle
\setlength{\parindent}{2em} 
\begin{abstract}
In this work, we re-examined the ancient complex metric in the recent quantum picture of black holes as Bose-Einstein condensates of gravitons. 
Both black holes and particles can be described by the complex Kerr-Newman metric in a 6-D complex space, which appears as a 4-D spacetime for a real or imaginary observer because of the barrier of the horizon.
As two kind of complex black holes, particle and black hole are complex conjugated and can convert into each other through a phase transition.
From the view of an observer in 3-D real space, an elementary particle with spin appears as an imaginary black hole in an anti-de Sitter space. 
The self-gravitational interaction of a particle as an imaginary black hole makes it obtain its wave-like nature in 4-D spacetime.
\end{abstract}


\section{Introduction}
The concern over the potential link between the black hole and the particle has a long and continuous history because it may provide us useful information about the connection between general relativity and quantum mechanics. 
In 1935, trying to in search of a geometric model for elementary particles, Einstein and Rosen \cite{einstein_particle_1935} provided a speculative structure now known as the Einstein–Rosen bridge. 
In 1968, Carter \cite{carter_global_1968} found that the Kerr–Newman solution \cite{newman_metric_1965} has a gyromagnetic ratio g=2 like the Dirac electron. 
Then, the Kerr–Newman electron has received constant attention \cite{debney_solutions_1969,israel_source_1970,d._ivanenko_gravitational_1975,barut_zitterbewegung_1981,lopez_extended_1984,lopez_internal_1992,burinskii_string-like_1993,israelit_classical_1995,finster_particlelike_1999,burinskii_super-kerr-newman_1999,burinskii_gravitational_2003,arcos_kerrnewman_2004,burinskii_dirac-kerr-newman_2008,burinskii_source_2016,burinskii_new_2017}
 and obtained supports from string theory \cite{holzhey_black_1992,sen_rotating_1992,sen_extremal_1995,nishino_stationary_1995,horowitz_rotating_1996}. 
 What’s more, there also have been suggestions that black holes should be treated as elementary particles \cite{t_hooft_black_1990,hawking_gravitational_1971,susskind_speculations_1993,susskind_black_1994,russo_asymptotic_1995,duff_new_1994,duff_massive_1995,sen_black_1995,hull_unity_1995,townsend_eleven-dimensional_1995,witten_string_1995,strominger_massless_1995,greene_black_1995}.
 
 Complex metric, provided by Newman and his co-workers in their derivation of the Kerr-Newman metric \cite{newman_metric_1965}, has been found to be a useful mathematical tool in various problems \cite{newman_maxwells_1973,gibbons_cosmological_1977,brown_complex_1991,burinskii_kerr_1998,burinskii_kerr-schild_2000,newman_classical_2002,burinskii_complex_2003}. Recently, a quantum picture of black holes as Bose-Einstein condensates(BEC) of gravitons 
 \cite{dvali_black_2013-3}. 
 In this picture, we found that complex Kerr-Newman metric has a deep physical meaning rather than just a mathematical model. 
 In a 6-D complex space, both common black holes and elementary particles are found to be special cases of the complex Kerr-Newman black holes, which can turn into each other through a phase transition. 
 By analysing the metric of a particle in the imaginary space, we found that the wave-like nature of a particle in 4-D spacetime is a result of the self-gravitational interaction of a particle as a black hole in the imaginary space.

 \section{Phase transition of complex black hole}
 
 \subsection{Common black hole as a  particular solution of complex black hole}
 The Kerr-Newman metric describes a general black hole with both charge and spin \cite{newman_metric_1965}. 
 The radius of its two horizons ($r_\pm$) are
 \begin{equation}
{r_ \pm } = m \pm \sqrt {{m^2} - {a^2} - {Q^2}} 
\label{lab1}
\end{equation}
where $m$ is its mass, $a$ is its angular momentum per unit mass, and $Q$ is its charge,$c = \hbar = G = {k_B} =1$ is used in this work ($c$ will appear where the speed of light needs to be stressed).
Equation (\ref{lab1}) seems to lose its physical meaning when $m^2< a^2+Q^2$. 
However, if a horizon can have complex radius, the physical meaning of this equation can be further expanded.
Re-writing equation (\ref{lab1}), we can obtain
\begin{equation}
{r_ \pm } = m \pm i\sqrt {{a^2} + {Q^2} - {m^2}} 
\label{lab2}
\end{equation}
The real radius of the complex horizon ($r_R$) is
\begin{equation}
r_R=m 
\label{lab3}
\end{equation}
In the 3-D real space, an elementary particle will appear as 0-D point in low energy if it can be described by equation (\ref{lab2}) because its $r_R$ is much smaller than Planck length and too small to be measured by current technology. This agrees with standard model. 
The imaginary radius of the complex horizon ($r_I$) is
\begin{equation}
{r_I} =  \pm i\sqrt {{a^2} + {Q^2} - {m^2}} 
\label{lab4}
\end{equation}

\begin{figure}[htbp]
	\centering
	\includegraphics[scale=0.4]{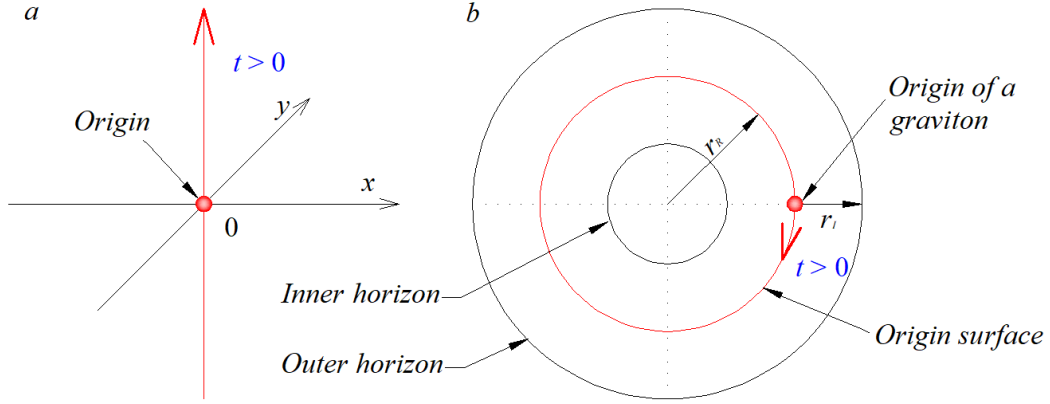}
	\caption{\textbf{Phase transition of complex black hole.} 
		After the point of phase transition, $r_I=0$, the imaginary radius is realized and the particle change into a black hole. 
		The objects in $a$ are corresponding to the objects in $b$: 0-D origin point vs 2-D origin surface, 1-D time dimension vs 3-D imaginary space. 
		The black holes is Bose-Einstein condensates of $N$ gravitons. For any graviton, its origin is a point on the origin surface.
		}
	\label{fig:fig1}
\end{figure}

With the increase of its $m$, $r_I$ of a particle reduces continuously to 0$i$ and then be realized, which means that the particle as a complex black hole changes into a common black hole. 
The phase transition point of the complex black hole is an extreme black hole, $r_I=0$. At the same time, $r_R$ of a particle increases continuously. 
$r_R$ not only characterizes the size of the particle, but also defines the boundaries of the 3-D real space for other observers.
Therefore, in the rest frame of the particle, the increasing $r_R$  can be understood as an expansion of the coordinate origin from a 0-D point to a 2-D spherical surface with radius of $r_R$ (as shown in Fig.\ref{fig:fig1}).

In this way, the inner space of a common black hole bordered by its inner and outer horizons is in fact a realized imaginary space embedded in the 3-D real space (while the space within its inner horizon is imaginary space). 
All the points in this realized imaginary space share the same real radius although their imaginary coordinates can be different.  
If we consider the rotational symmetry, these points can be considered indistinguishable points in the real space. 
This agrees with quantum picture of black holes as Bose-Einstein condensates of $N$ gravitons\cite{dvali_black_2013-3}. 

\subsection{Gravitons' motions in a BEC black hole}

The origin surface of a complex black holes, which is one of the most important physical contents in this work, seems counter-intuitive.  
In the  quantum picture of black holes as Bose-Einstein condensates of $N$ gravitons\cite{dvali_black_2013-3}, this concept is easier to understand.

According to \cite{dvali_black_2013-3}, the number of gravitons of a Schwarzschild black hole in BEC with mass of $M$ is 
\begin{equation}
	N=M^2
	\label{lab5}
\end{equation}
The mass of every graviton is
\begin{equation}
	m_g=1/M
	\label{lab6}
\end{equation}
The origin surface with many points can be regarded as a collection of the origins of $N$ gravitons (as shown in Fig.\ref{fig:fig1}).

The de Broglie wavelength of any graviton 
\begin{equation}
	\lambda  = \frac{{2\pi }}{{{m_g}}} = 2\pi M
	\label{lab7}
\end{equation}
is found to be the circumference of a  circle with radius of $r_I$ ($=M$ for this Schwarzschild black hole with mass of $M$), which implies that a graviton of the  Schwarzschild black hole in BEC may be a stand wave  centered on its origin on the origin surface.
In fact, a graviton without rest mass moves at the speed of light. 
If it does a uniform circular motion with a radius of $r_I$ around its origin, it will complete one cycle in one period.

In the 4-D spacetime, the origin of a particle’s rest frame moves along the time dimension at the speed of light ($a$ of Fig.\ref{fig:fig1}). 
After the phase transition of space, the time dimension is unfolded into the 3-D imaginary space.
The moving distance of the origin in the time dimension will be a large value after a long time, which seems an impossible motion in the limited realized imaginary space.
A reasonable solution is that the motion of the origin of each graviton is an uniform circular motion on the origin surface at the speed of light. 
The clockwise and counterclockwise rotation correspond to $t>0$ and $t<0$, respectively.
In this way, the synthetic movement of the graviton's 
circular motion and the motion of its origin will be a uniform circular motion on the horizon with speed of light.

In this subsection, combining complex metrics \cite{newman_metric_1965}  and the quantum black holes as  BECs of gravitons 
\cite{dvali_black_2013-3}, we provide a possible picture of how gravitons move in a black hole.
This picture is found to be an key to understand the geometric origin of the de Broglie waves (will be discussed in detail  later).

\section{Particle as imaginary black hole in AdS}

\subsection{Singularity as the origin of time}
What kind of geometry does a particle have in the hidden 3-D imaginary space?
Penrose's idea about singularity \cite{penrose_gravitational_1965} provides us useful clue.

 According to Penrose \cite{penrose_gravitational_1965}, the singularity is the origin of time. 
In the 6-D complex space, the counterpart of the origin of time in 4-D spacetime is the origin of the 3-D imaginary space. 
The singularity of a common Kerr–Newman black hole appear as a ring on its equatorial plane with a radius of 
\begin{equation}
{r_{s,R}} = a
\label{lab8}
\end{equation}
Any direction of rotation is mathematically equivalent because of the rotational symmetry of the 3-D real space. 
Therefore, the ring singularity can be regarded as a special solution of a sphere singularity after the direction of rotary axis is locked. 
The singularity of a common real black hole is completely enclosed by the event horizon.
From the above section, we know the space within the event horizon is in fact a imaginary space. 
Therefore, the origin surface should have an imaginary radius with a modulus of $a$.
In this way, the origin surface of the particle in imaginary space has a radius of
\begin{equation}
{r_0} = ia
\label{lab9}
\end{equation}
and the horizon of a particle in the 3-D imaginary space has a radius of $r_I$ (as shown in $I of Fig.2$), which means that a particle appears as an imaginary black hole, the horizon radii of which are
\begin{equation}
{r_ \pm } = ia \pm i\sqrt {{a^2} + {Q^2} - {m^2}} 
\label{lab10}
\end{equation}
Equation (\ref{lab10}) can be rearranged as 
\begin{equation}
{r_ \pm } = ia \pm \sqrt {{{(ia)}^2} - {{(im)}^2} - {Q^2}} 
\label{lab11}
\end{equation}
Comparing equations (\ref{lab1}) and (\ref{lab11}), we can obtain the equivalent mass, angular momentum per unit mass of the imaginary black hole of the particle as
\begin{equation}
\left\{ {\begin{array}{*{20}{c}}
{{M_i} = ia}\\
{{a_i} = im}
\end{array}} \right.
\label{lab12}
\end{equation}

\subsection{Hawking temperature of a particle as imaginary black hole}

The imaginary black hole has a Hawking temperature of
\begin{equation}
{T_i} = \frac{1}{{2\pi }}\frac{{{r_ + } - {r_ - }}}{{2(r_ + ^2 + a_i^2)}}
\label{lab13}
\end{equation}
which has an imaginary value.

A black hole can harvest energy from its environment and lose energy through Hawking radiation. 
Therefore, energy balance is a necessary condition for a stable black hole. 
The Hawking temperature of a particle as an imaginary black hole is therefore a good mark of the energy level of its local imaginary space. 
According to the work of Deser and Levin \cite{deser_accelerated_1997}, an inertial observer in a de Sitter ($dS$) or anti-de Sitter ($AdS$) spaces with cosmological constant $\Lambda$ will measure a temperature of
\begin{equation}
{T_\Lambda } = \frac{1}{{2\pi }}\sqrt {\frac{\Lambda }{3}} 
\label{lab14}
\end{equation}
When $\Lambda<0$, this temperature will have an imaginary value. 
Therefore, from the view of an observer in the real space, the imaginary space of our universe is an $AdS$ space. 
According to the symmetry of complex space, the time dimension of the $AdS$ space is folded from the 3-D real space.

It should be specially stated that the imaginary values of physical concepts in the imaginary space including mass and length  are only relative to the observers in the real space.
From the view of any observer in the imaginary space, the imaginary black hole is just a common real one and the imaginary space is a $dS$ space. 

\subsection{Evolution of a complex black hole}

The presence of complex black holes (including common black holes and particles) makes the coordinate origin of the complex space expands from the point of $0+0i$ to a complex spherical surface with a radius of
\begin{equation}
{R_0} = m + ai
\label{lab15}
\end{equation}
The imaginary black hole of a particle also have a ring singularity with a radius of 
\begin{equation}
{r_{s,I}} = a_i=im
\label{lab16}
\end{equation}
The modulus of $r_{s,I}$ equals to the radius of its origin surface in the real space.
The ring singularity of a common real black hole also have this characteristic.
Therefore, the ring singularity of a real or imaginary black hole is in fact a section
of the origin surface in its complex conjugate space after the  direction of rotary axis is locked. In the complex space, singularity is covered by horizon.

\begin{figure}[htbp]
  \centering
  \includegraphics[scale=0.4]{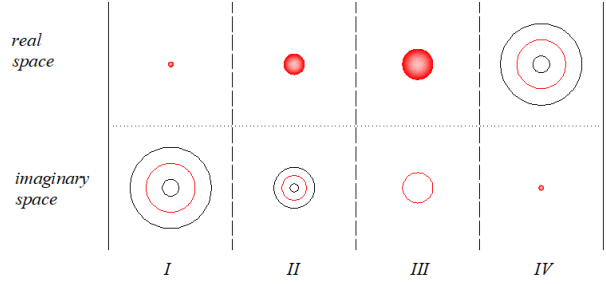}
  \caption{\textbf{Evolution of a complex black hole.} $I$: the particle appears as a point-like particle in the real space while as an imaginary black hole in the imaginary space; $II$: the point expands while the imaginary black hole shrinks; $III$: extreme black hole and extreme imaginary black hole; $IV$: real black hole and imaginary point-like particle. Red circles or spherical surfaces represent origin surfaces, while the black circles represent inner or outer horizon.}
  \label{fig:fig2}
\end{figure}

The evolution of a complex black hole in complex space is shown in Fig.(\ref{fig:fig2}). 
We can found that two special cases of complex black holes, particles and common black holes, are complex conjugated.

\section{Geometric origin of de Broglie wave}

\subsection{Wave-like nature as a result of gravitons' motion}

For all elementary particles except Higgs boson in the standard model, the following equation
\begin{equation}
{r_I} = i\sqrt {{a^2} + {Q^2} - {m^2}}  \approx ia
\label{lab17}
\end{equation}
is a sufficiently accurate approximation. Therefore, the imaginary Kerr-Newman black hole of the particle is an approximate imaginary Schwarzschild black hole. 
According to \cite{dvali_black_2013-3}, the imaginary component of the mass of every graviton of the imaginary black hole in Bose-Einstein condensate is
\begin{equation}
{m_{g - i}} = \frac{i}{{\left| {{M_i}} \right|}} = \frac{i}{a}
\label{lab18}
\end{equation}
while the number of the gravitons is
\begin{equation}
N = {\left| {{M_i}} \right|^2} = {a^2}
\label{lab19}
\end{equation}

From the view of an imaginary stationary observer in the rest frame of the particle, the motions of the gravitons of its imaginary black hole in BEC are the synthesis of the motions of their origins and their circular motions around their origins.
When the 3-D imaginary space folds to the 1-D time dimension of the 4-D spacetime, these circular motions make particles obtain their wave-like nature. 
The real part of the complex wave function is the component in a certain direction of the 3-D imaginary space which acts as the time dimension of the 4-D spacetime, while the imaginary part is the component perpendicular to this direction.
In the following, we will derive the geometric origin of the plane wave of a free Dirac fermion ($L=1/2$). The question will be discussed in the free particle's inertial coordinate frame.

\begin{figure}[h]
	\centering
	\includegraphics[scale=0.4]{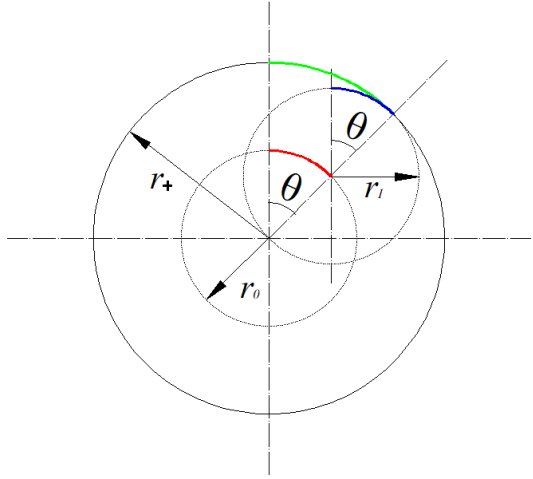}
	\caption{\textbf{Geometric origin of wave nature of a Dirac fermion.} 
		The motions of the gravitons of the imaginary black hole in Bose-Einstein condensate make the particles obtain their wave-like nature. 
		The red, blue, and green bold short arcs 
		represent the displacement of a graviton's origin,
		the displacement caused by the circular motion around its origin, and its total displacement, respectively.
	}
	\label{fig:fig3}
\end{figure}

For the imaginary black hole of a free Dirac fermion, we assume that two circular motions of its gravitons are in one plane (Fig.\ref{fig:fig3}).
During a time interval of $0.5t_0$, the displacement of its origin ($\Delta l_{i-0}$, the red bold short arc in  Fig.\ref{fig:fig3}) is
\begin{equation}
	\Delta {l_{i - 0}} = ic \times 0.5t_0
	\label{lab20}
\end{equation}
During the same time, the displacement caused by the circular motion around its origin ($\Delta l_{i-g}$, the blue bold short arc in Fig.\ref{fig:fig3}) is
\begin{equation}
\Delta {l_{i - g}} = ic \times 0.5t_0
	\label{lab21}
\end{equation}
The phase angle of the graviton, $\theta$, will be
\begin{equation}
	\theta = \frac{{\Delta {l_{i - 0}}}}{{{r_0}}} = \frac{{\Delta {l_{i - g}}}}{{{r_I}}}
	\label{lab22}
\end{equation}
In the free particle's inertial coordinate frame, the particle is stationary at the coordinate origin, which means that $v=0$. Therefore,
\begin{equation}
{r_0}(v = 0) = {r_I}(v = 0) = \frac{iL}{m_0}
	\label{lab23}
\end{equation}
where $m_0$ is the rest mass of the particle.
Substituting $L=1/2$ and equation (\ref{lab23}) into  equation (\ref{lab22}) yields
\begin{equation}
\theta =m_0ct_0
\label{lab24}
\end{equation}

From the view of the observer in the 3-D imaginary space, the resultant motion of the graviton happen on the outer horizon of the imaginary black hole, a spherical surface with a radius of $r_+$($\approx 2ia$). 
The total displacement of the graviton ($\Delta l_i$, the green bold short arc in Fig.\ref{fig:fig3}) is
\begin{equation}
\Delta {l_i} = \theta \times r_+= ict_0
\label{lab25}
\end{equation}

The uniform energy of every graviton of the imaginary black hole in BEC means that the phase difference between them remains the same.
All the original positions of their origins are components of the starting point  of the particle in time dimension. 
Therefore, the phase angle of the particle is $\theta$ described in equations (\ref{lab23}). 
The clockwise and counterclockwise rotation of the gravitons on the horizon of their imaginary black hole correspond to the two sign of the spin, respectively.

\subsection{Lorentz transformation as a conformal transformation}

The phase angle of the wave function of a stationary particle ($v=0$) is given above. Substituting the Lorentz transformation
\begin{equation}
{t_0} = \gamma (t - \frac{{vx}}{{{c^2}}})
\label{lab26}
\end{equation}
where $\gamma$ is the Lorentz factor
\begin{equation}
\gamma = \frac{c}{{\sqrt {{c^2} - {v^2}} }}
\label{lab27}
\end{equation}
into equation (\ref{lab24}), we can get the phase angle of a free particle with speed of $v$ as
\begin{equation}
\theta ' = mct - px
\label{lab28}
\end{equation}

\begin{figure}[htbp]
	\centering
	\includegraphics[scale=0.6]{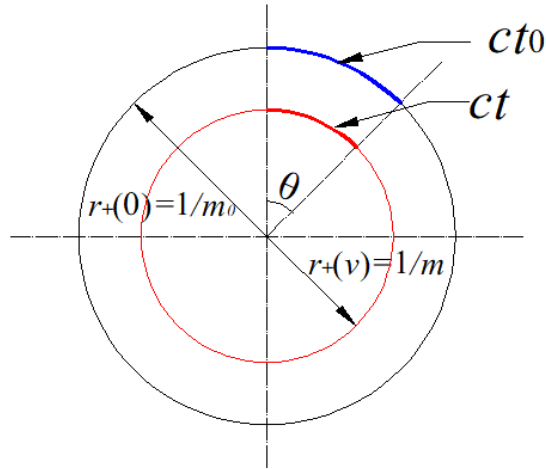}
	\caption{\textbf{de Broglie’s internal clock.}  
	 When a particle gets more energy, its imaginary black hole  shrinks but the phase angle does not change,
	 which will result in time dilation if the speed of light is a constant.
	}
	\label{fig:fig4}
\end{figure}

From equation (\ref{lab28}), we can get some interesting information about the motion of the particle as a black hole in the imaginary space.
First, when $x=0$,
\begin{equation}
\theta ' = mct = m_0ct_0 = \theta
\label{lab29}
\end{equation}
which means that the phase angle doesn't change when the imaginary black hole  shrinks (as shown in Fig.\ref{fig:fig4}, which gives a visual display of time dilation in special relativity).
Therefore, Lorentz transformation is a conformal transformation for observers in the imaginary space.

Being an imaginary black hole as BEC of gravitons, a particle will have a internal clock as conjectured by de Broglie because of the motions of its gravitons.
This internal clock is hidden in the imaginary space.

\begin{figure}[h]
	\centering
	\includegraphics[scale=0.5]{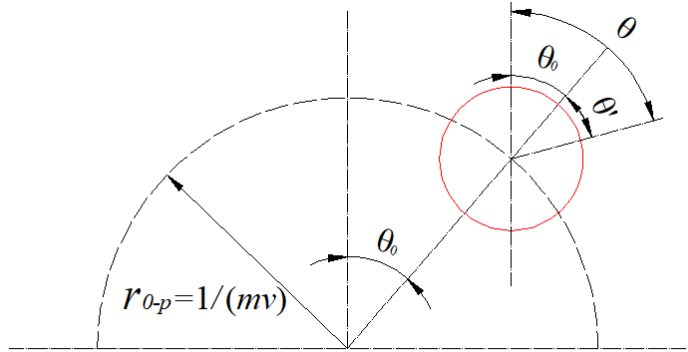}
	\caption{\textbf{Contribution of momentum to phase angle.}  
	The imaginary black hole of a particle is doing an uniform circular motion on an origin surface with radius of $1/mv$, which makes different points in the 3-D real space  share an equivalent imaginary coordinate.
	In this way, a particle can appear at these points simultaneously.
	}
	\label{fig:fig5}
\end{figure}

Then, let us analyze the physical meaning of the term of "$-px$" in equation (\ref{lab28}). 
As a matter wave, a particle can appear in different locations of the 3-D real space with  a certain probabilities simultaneously.
From section 2, we know that all the gravitons of a black hole in BEC share an equivalent real coordinate.
If the points in the 3-D real space, at which the particles can appear at the same time, share an equivalent imaginary coordinate, the
wave nature will be a natural result. 
From the view of any graviton of a particle, the uniform linear motion of the particle with a speed of $v$ in the 3-D real space appears as the motion of its imaginary black's origin with the speed of $iv$ in the 3-D imaginary space. 
We assumed that the imaginary black hole is doing an uniform circular motion on an origin surface with radius of
\begin{equation}
{r_{0-p}} = \frac{i}{p}
\label{lab30}
\end{equation}
where $p=mv$ is the momentum of the particle (as shown in Fig.\ref{fig:fig5}). 
In this way, we can get the contribution of a particle's momentum to phase angle.
The phase angle caused  by the motion of the imaginary black hole on the origin surface with a speed of $iv$ is 
\begin{equation}
{\theta _0} = \frac{{ivt}}{{{r_{0-p}}}}
\label{lab31}
\end{equation}
The final net phase angle is
\begin{equation}
\theta ' = \theta  - {\theta _0} = mct - mvt
\label{lab32}
\end{equation}

\subsection{Negative energy solutions of Dirac equation}
\begin{figure}[h]
	\centering
	\includegraphics[scale=0.6]{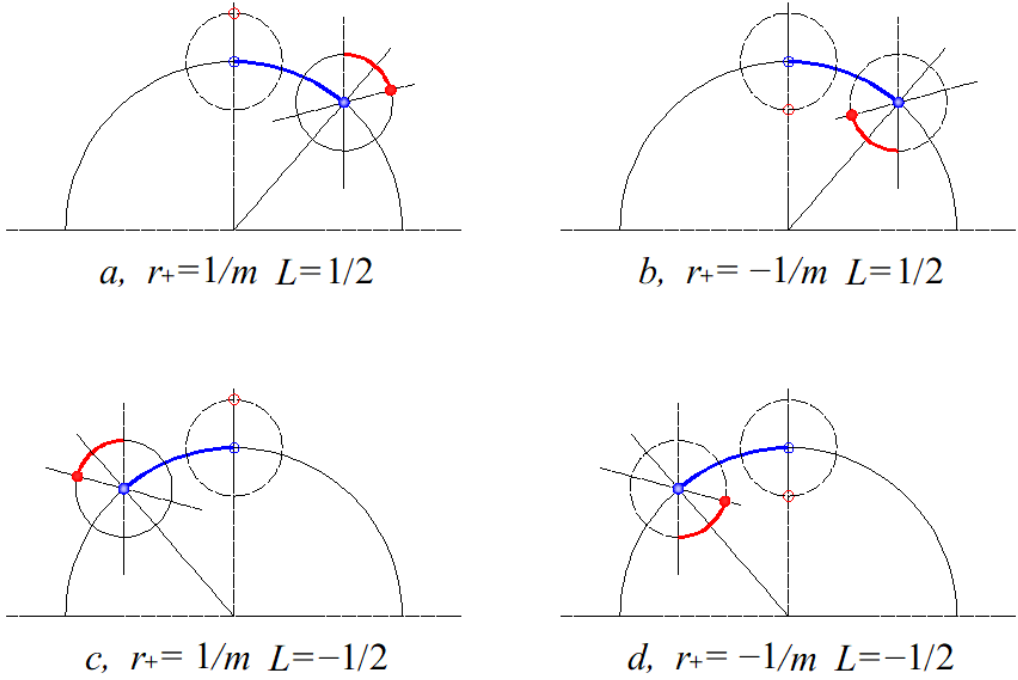}
	\caption{\textbf{Four solutions of Dirac equation.}  
	Blue arcs indicate the displacement of the imaginary black hole as Bose Einstein condensate of gravitons, while red arcs indicate the displacement of one of its gravitons (hollow circle indicates the starting positions and a solid circle indicates the current positions).
	}
	\label{fig:fig6}
\end{figure}

Since its origin  is on the origin surface with radius of $r_{0-p}$, an imaginary black hole with a negative radius of "$-r_+$" still has physical meaning.
These possible states are found to correspond to the negative solutions of Dirac equation, four solutions of which are shown in Fig.\ref{fig:fig6} .

Comparing the motion of a graviton in a positive  black hole in BEC ($a$ of Fig.\ref{fig:fig6}) and that of a negative energy black hole in BEC ($b$ of Fig.\ref{fig:fig6}), we can find that there is a central symmetry between them  with respect to the center of the imaginary black hole.
Similarly, there is a mirror symmetry between two spin states of the particle
($a$ vs $c$, $b$ vs $d$).

In a space where the speed of particle is smaller than that of light ($v<c$), the following relationship can always be satisfied
\begin{equation}
{r_{0 - p}} > \left| {{r_ + }} \right|
\label{lab33}
\end{equation}
Therefore,  any relativistic wave equation that satisfies the Lorentz transformation will have negative energy solutions.

\section{Conclusion and Discussion}
\subsection{Conclusion}
Thanks to the recent quantum picture of black holes as Bose-Einstein condensates of gravitons \cite{dvali_black_2013-3}, we are able to re-examin the "ancient" complex metric \cite{newman_metric_1965} in this work. 
The conjugate symmetry between a common black hole and elementary particle in a 6-D complex space is found. 
For any observer in the 3-D real (or imaginary) part of the complex space, he (or she) can’t observe the imaginary (or real) space directly because of the barrier of the horizon.
The three imaginary (or real) dimensions will fold into a time dimension, which makes the 6-D complex space appears as a 4-D spacetime.

An elementary particle with spin appears as an imaginary black hole with a mass of $ia$ in an AdS space ($a$ is the spin pure unit mass).
In the quantum picture of black hole, this imaginary black hole consists of $N=a^2$ gravitons. 
The motions of these component gravitons make the particle they make up obtain its wave-like nature. 
Therefore, the quantum properties of the particles we observed in a 4-D spacetime is a result of the gravitational effect in a 3-D imaginary space. 
With a time dimension folded from the 3-D real space, the 4-D imaginary spacetime is found to be a AdS space for a real observer. 
This agrees with the AdS/CFT correspondence proposed by Maldacena \cite{maldacena_large-n_1999}.

\subsection{Discussion}

This work can provide us a new perspective to understand some problems in quantum mechanics and general relativity such as black hole information paradox.
The black hole information paradox shows the conflict between quantum mechanics and general relativity. 
An important recent development in this area is AMPS firewall \cite{almheiri_black_2013}. 
In order to resolve the AMPS firewall paradox, Maldacena and Susskind \cite{maldacena_cool_2013} provided the ER=EPR hypothesis that entangled objects may be connected through the interior via a wormhole, or Einstein-Rosen bridge. 
In the picture provide in this work, we can found that the creation of an entangled particle pair is indeed a creation of entangled black hole as conjectured in ER=EPR. 

In addition, according to Hawking's picture about black hole's Hawking radiation \cite{hawking_black_1974}, we know that when a particle with positive energy escapes from the horizon, its antiparticle with negative energy will fall to the singularity. 
The singularity of the imaginary black hole is the origin of the real space.
Therefore, if these associated negative particles of the Hawking radiation of the imaginary black hole cross its singularity rather than end at it, the clouds of virtual particles around a particle in the real space will have a gravitational origin.








\section*{Acknowledgements}

 This work is supported by the National Science Foundation of China (No. 51736004 and No.51776079).

\bibliographystyle{unsrt}  
\bibliography{references}  


\end{spacing}

\end{document}